# Development of In Situ Acoustic Instruments for The Aquatic Environment Study

**A.N. Grekov[*], N.A. Grekov, E.N. Sychov, K.A. Kuzmin**

Institute of Natural and Technical Systems,
Russia, Sevastopol, Lenin St., 28
*E-mail: oceanmhi@ya.ru

Based on the analysis of existing acoustic methods and instruments, a prototype of an automated instrument has been developed to perform joint measurements in situ of two parameters: sound speed and ultrasound attenuation. The device is based on existing sound velocity profilers. It was proposed to replace the TDC-GP22 converters used in the sound speed meter ISZ-1 with more advanced modern modified converters TDC-GP30, which can significantly improve the accuracy of measuring the amplitude of the reflected acoustic signal. The programs for processing signals from the primary acoustic transducer have been developed. The model of the device passed preliminary tests.

**Introduction**. The great variability of the physicochemical and biological characteristics of inland water bodies and shelf zones of the seas sets the task of operational control of the dynamics of the processes under study and the improvement of environmental monitoring systems. The creation of in situ acoustic systems, as well as the development of their methodological support, are priority directions in the development of marine instrumentation.

In order for acoustic meters of ultrasound velocity and attenuation to be in demand in oceanographic research, they must be reliable, not expensive to manufacture and operate, have a fairly simple calibration procedure and the necessary set of application methods, including equations and models.

It is known that the advantage of the acoustic method for measuring the velocity and absorption of ultrasound is that the transparency of a liquid does not matter, in contrast to the optical method, where only optically transparent liquids are investigated.

For low-viscosity liquids, which include seawater, the upper frequency range of acoustic waves in which the speed of sound C and the attenuation coefficient $C$ are currently studied reaches $1 \cdot 10^{10}$ Hz. Technically, to obtain such frequencies, thin piezoelectric and piezo-semiconductor films are used [1]. However, in real conditions, it is difficult to use frequencies higher than $1 \cdot 10^9$ Hz to study water, because measurements must be taken at distances of hundreds of microns and work with high intensity waves.

Despite the fact that the velocity and attenuation of ultrasound have been studied in detail for distilled water, and experimental results have been obtained on the dependence of the attenuation of ultrasound on frequency and temperature, similar data on seawater for various seas and oceans, especially shelf zones, are completely insufficient.

Theoretical calculations of the velocity of propagation and absorption of ultrasound in natural liquids based on the molecular theory have not yet been substantiated, since there is currently no rigorous theory [2–14].

However, researchers continue to accumulate experimental data on the propagation velocity and attenuation of ultrasound and, thus, are trying to determine the limits of applicability of empirically established laws and use them for practical purposes. New methods for calculating the acoustic properties of natural waters are being developed, and their relationship with thermodynamic characteristics is being established, which in some cases has some advantages over traditional measurement models, for example, by analyzing the aquatic environment using samples taken with bathometers.

To create automated devices that measure the absorption of ultrasound in natural waters and operate in situ, it is necessary to assess the capabilities of the technologies achieved. Meters of the speed of sound, the analysis of which is given in [15], are widely known in oceanographic work. However, in the future, such devices can be supplemented with an ultrasound absorption meter without significant financial costs.

**Main part.** Consider water sound speed meters designed and built by various manufacturers. All these devices, after retrofitting them with modern microcircuits and software, including the methods of their use, can also be used to measure the attenuation of an acoustic signal in water.

One of the leading manufacturers of sound velocity profilers (SVP) in water is the English company Valeport Limited (Great Britain). All profilers of this company are supplied with the same type of sound velocity sensors with a measuring range of 1400–1600 m / s. The measurement of the speed of sound is based on a time-of-flight measurement. Each sound velocity measurement is performed with one acoustic pulse passing through a stable measurement base. Depending on the length of the measuring base, which is formed by highly stable rods made of carbon composite material, different resolution and accuracy in terms of sound velocity can be obtained. Technical and metrological data of sound speed sensors at a piezoceramic radiation frequency of 2.5 MHz (Valeport Limited) are shown in Table 1.

Table 1. Technical and metrological characteristics of sensors

| Sound speed sensor, mm | Sensitivity, m/s | Range, m/s | Signal stability, m/s | Calibration accuracy, m/s | Total error, m/s | Frequency, Hz |
|---|---|---|---|---|---|---|
| 25 | 0,001 | 1400÷1600 | ±0,01 | ±0,085 | ±0,095 | 22 |
| 50 | 0,001 | 1400÷1600 | ±0,006 | ±0,054 | ±0,06 | 16 |
| 100 | 0,001 | 1400÷1600 | ±0,003 | ±0,027 | ±0,03 | 11 |

The Valeport sound velocity profilers are integrated into the MIDAS series and can be included in both sounding and buoy measuring systems.

Due to the design features of the primary transducer, all measuring channels of the sound velocity of Valeport devices have a disadvantage associated with the dependence of the measuring base on pressure.

The Institute of Natural and Technical Systems (IPTS) has developed its own SVP ISZ-1 [16]. Its technical characteristics are presented in Table 2.

Consider the technical capabilities of the ISZ-1 sound velocity measuring channel:
- the sampling frequency of the measuring channel is 18 Hz;
- the length of the measuring base of the sound velocity sensor is 6 cm;
- error in measuring the speed of sound in static mode 0.02 m/s;
- the frequency of the radiated radio pulses is 2 MHz

The spatial scale of the sound velocity sensor is determined by the measuring base and the diameter of the piezoceramic, in our case l = 6 cm, d = 1 cm.

Table 2. Technical characteristics of the ISZ-1 probe

| Measurable and design parameters | Range measurements | Random error | Error |
|---|---|---|---|
| Sound speed, m/s | 1375 ÷ 1900 | 0,001 | ± 0,02 |
| Water temperature, °C | -2 ÷ +35 | 0,001 | ± 0,01 |
| Hydrostatic pressure, kPa | 0 ÷ 20000 | 2 | ± 20 |

The design of the sound velocity sensor of the ISZ-1 device is made in such a way that its measuring base does not depend on external pressure and excludes additional pressure calibration, which improves the dynamic characteristics. The device can work up to a depth of 6000 m with the installation of a suitable pressure sensor.

The sound speed channel is calibrated under laboratory conditions using the equation for the speed of sound in seawater [17]. Del Grosso also published an equation for the speed of sound in distilled water, together with Mader [18]. Distilled water has a clear advantage over seawater in that fewer parameters need to be controlled, namely pressure and temperature. Therefore, at a fixed pressure (i.e., atmospheric, in laboratory conditions), it is quite easy to accurately control distilled water in terms of temperature. Thus, the authors of [19] estimate that the equation of Del Grosso and Madera has an accuracy index equal to ± 0.015 m/s; this is significantly better than the sea water equations.

It is known that the existing sound speed meters operate at frequencies of several MHz [15]. The device ISZ-1 developed by us operates at a frequency of 2.0 MHz

The first attempt to determine the ultrasound attenuation was made on the existing ISZ-1 device, where the TDC-GP22 converter was installed. This converter uses the First Wave Mode [20] to determine the signal attenuation.

The measured attenuation in this mode is an additional result when implementing the (Zero-Crossing Detection) method [21], which is used to improve accuracy and reduce the influence of noise in determining time intervals.

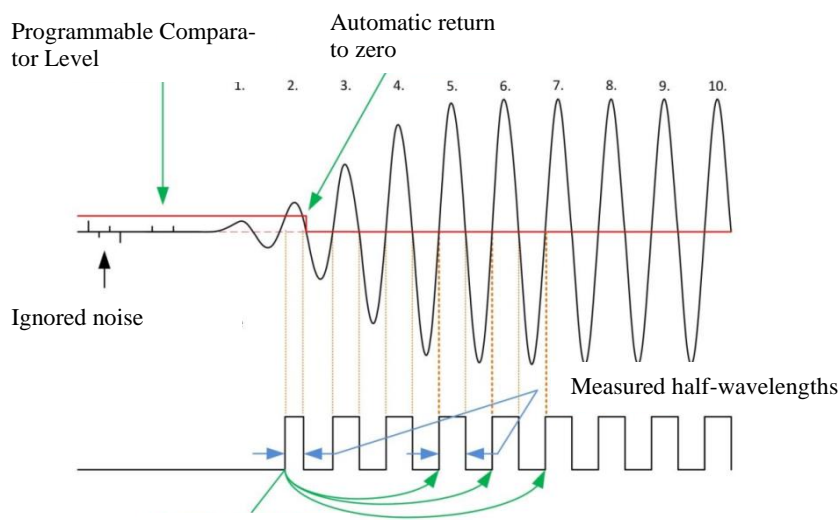

**Fig. 1.** Receiving signal, comparator threshold level and pulses arriving time-to-digital converter

Figure 1 illustrates this mode of operation graphically. After the signal is emitted into the environment, the receiver of the converter sets the comparator operation threshold to a programmable level (from 0 to 35 mV in 1 mV steps). After the level of the half-wave of the receiving signal reaches the set threshold, the converter measures the time interval between the rising and falling edge of this first half-wave and immediately resets the comparator threshold to zero. Then the converter measures the half-wave of one of the next three periods of the signal. This value is used as a reference. The ratio between the first half-wave and the reference (in the range from 0 to 1) is converted to the ultrasound attenuation.

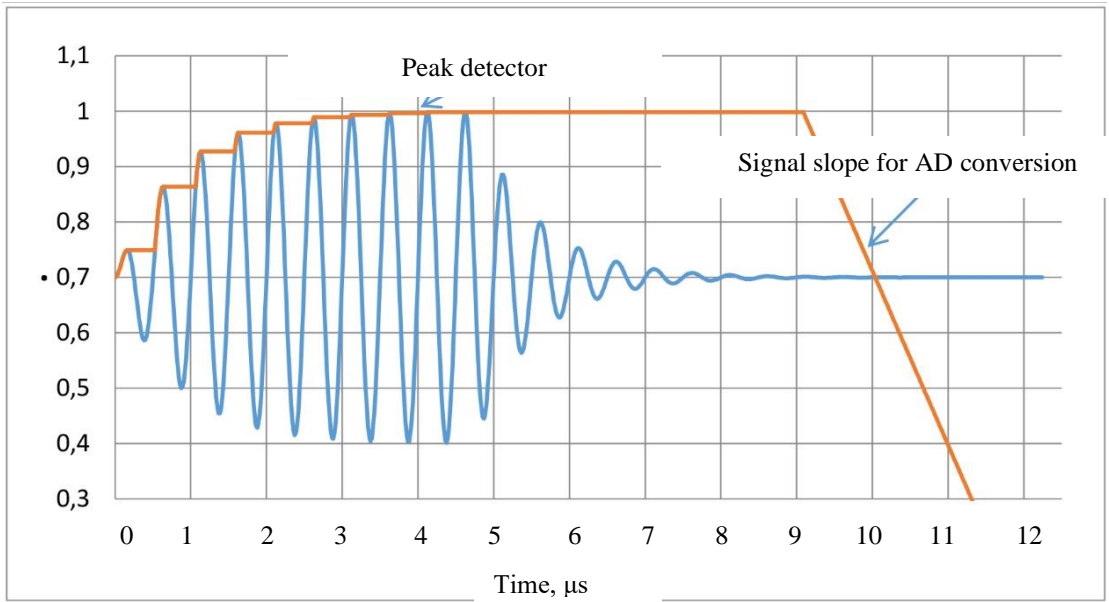

**Fig. 2.** Acoustic signal amplitude measurement circuit

Real attempts to use the First Wave Mode converter did not give satisfactory results due to the large error in determining the attenuation. The error in determining the amplitude reached 20%, which is not acceptable for our studies.

In contrast to the TDC-GP22, the TDC-GP30 introduces a new function for measuring the amplitude of the receiving signal, which significantly increases the accuracy of its determination [22]. Figure 2 shows the circuit for measuring the signal amplitude.

The amplitude measurement algorithm works as follows. After the reflected signal is received by the peak detector, the peak amplitude is recorded. Then, after the end of the action of the packet of pulses of the receiving signal using an analog-to-digital converter, taking into account the slope of the pulse, the time of its discharge is determined within the amplitude from the maximum to the establishment of the level, in our case, 0.7.

Based on the time measurement results, the amplitude of the acoustic signal is calculated using the following formulas:

$$V_{Up}[mV] = AMC_{Gradient}[mV/ns] * AM_{Up}[ns] - AMC_{Offset}[mV],$$

$$V_{Down}[mV] = AMC_{Gradient}[mV/ns] * AM_{Down}[ns] - AMC_{Offset}[mV],$$

where

$$V_{CAL}(typ) = V_{REF}/2 = 350 \text{ mV}, V_{REF} = \text{internal reference},$$

$$AMC_{Gradient}[mV/ns] = V_{CAL}[mV] / (AMC_H[ns] - AMC_L[ns]),$$

$$AMC_{Offset}[mV] = (2*AMC_L[ns] - AMC_H[ns]) * AMC_{Gradient}[mV/ns].$$

The use of the TDC-GP30 converter to determine the amplitude of the decaying signal is quite satisfactory for the developers. The main contribution to the error in determining the attenuation coefficient is made by the primary transducer, in our case the sound speed sensor. As already noted, all SVP sound speed sensors of devices are of the same type and operate on the principle of measuring the time difference between the probing and reflected signal on an exemplary base.

The use of the probing signal and the reflected acoustic signal to determine the attenuation coefficient is associated with some difficulties. The reflected signal is partially absorbed by the reflector. Let's consider this issue in more detail, assuming that the sensor is in sea water.

The reflection coefficient at the interface between two media is determined by the known relationship:

$$k = \frac{z_b - z_0}{z_b + z_0}, \tag{1}$$

where $k$ is the reflection coefficient; $z_b$ – specific acoustic resistance of water; $z_0$ – acoustic impedance of the reflector.

In turn

$$z_b = \rho_b c_b, \tag{2}$$

where $\rho_b$ – density of water; $c_b$ – speed of sound in water.

Likewise

$$z_0 = \rho_0 c_0, \tag{3}$$

where $\rho_0$ and $c_0$ are the density and speed of sound in the reflector.

According to our estimates, for a sound velocity sensor, where the reflector is made of stainless steel or titanium at a temperature of 20 °C, the reflection coefficient is about $k=0,93$. But, the value of $z_b$ depends on temperature, pressure and the amount of salts. Consequently, the coefficient $k$ is also variable and depends on these parameters.

Parameter $z_b$ can be calculated from the readings of measurements of the speed of sound, temperature and pressure, using expressions for determining the density of distilled and seawater [23].

When measuring the attenuation coefficient, it is necessary to know the initial amplitude 4 and the amplitude of the reflected signal $U_0$ at a known distance, and then, using formula $\alpha = \frac{1}{L} \ln \frac{U_H}{U_0}$, the attenuation coefficient $\alpha$ is calculated. The issue of determining the attenuation coefficient $\alpha$ for imported devices, where only one reflector is used in the sound speed sensor, is not considered in this work.

The ISZ-1 sound speed sensor, in contrast to foreign devices, has two reflectors; therefore, it is not difficult to determine the amplitude $U_0$. The difference in the amplitude of the signals between the receiving signals from the two reflectors will characterize the value of the attenuation coefficient of ultrasound in the medium along the length of the base between the two reflectors.

But given that the first reflector has a semitransparent design, its reflection coefficient is taken into account when calibrating the device.

**Description of the program.** Let us consider the scheme of the developed program for determining the speed and attenuation of ultrasound for the TDC-GP30 converter installed in the device prototype (Fig. 3). When the device is powered on, the main microcontroller performs a number of settings, including the settings for the interfaces with the on-board device and with the ultrasonic transducer microcircuit, after which it enters the autonomous mode. At any time, the device can receive a command from the on-board device, which switches it to one of the operating modes: telemetric, autonomous, working with memory or working with settings.

In telemetric and autonomous mode, a working cycle is carried out, during which the initial ultrasonic conversions are set up and started, the intermediate result is read, additional settings are based on the intermediate results, and the final result is read; depending on the mode, the result is stored in the internal memory of the device or transmitted to the on-board unit. In the mode of working with the memory of the device, the main microcontroller can receive commands for reading or formatting the memory and perform the necessary actions corresponding to them. In the settings mode, the main microcontroller receives a write command followed by a data packet and writes them to its non-volatile memory or receives a read command and transfers the contents of the settings area stored in the non-volatile memory.

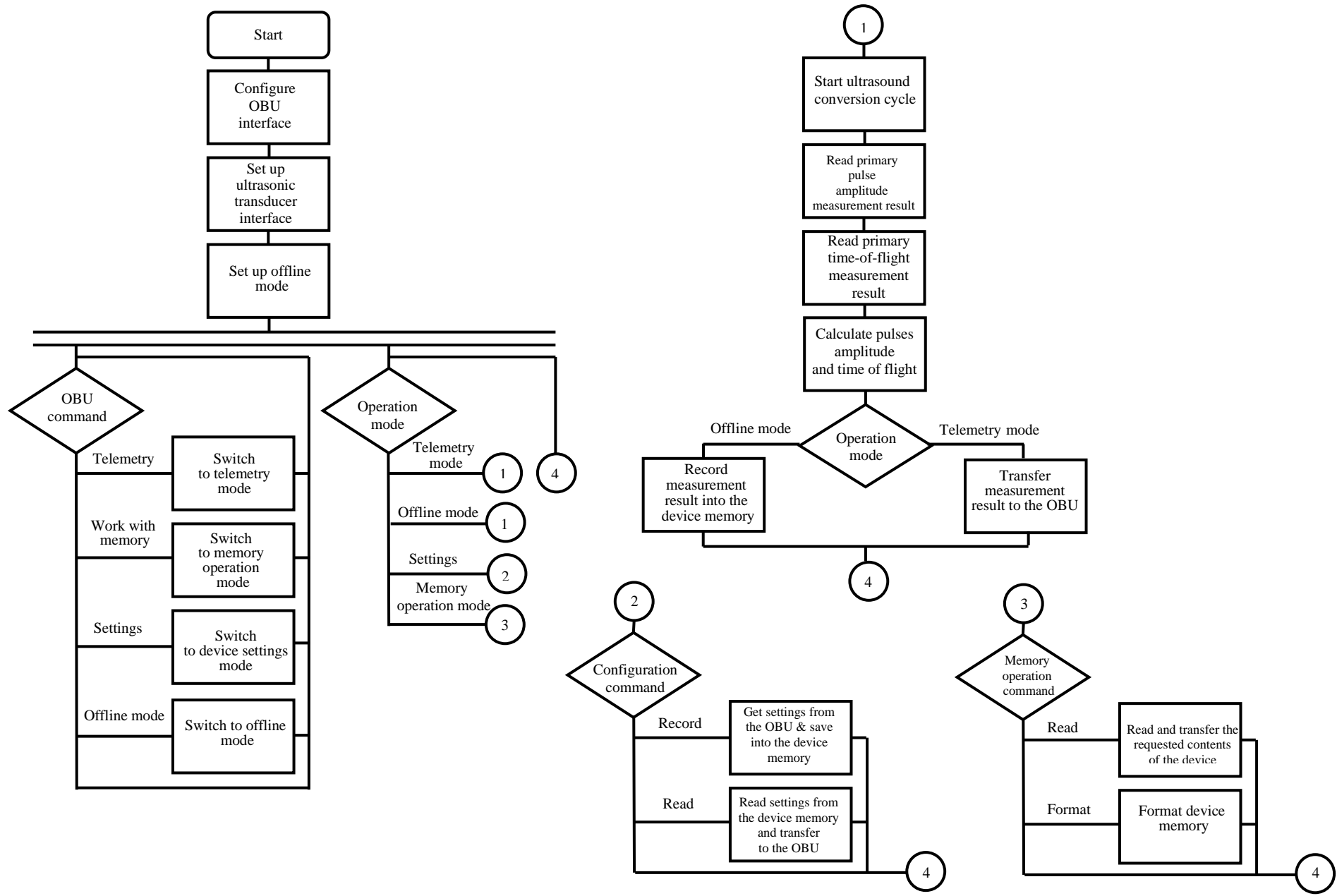

**Fig. 3.** Scheme of the program for calculating the time and amplitude of the ultrasonic signal

**Conclusion**. It is proposed to equip the sound velocity meters with an additional channel for determining the sound in water, which will expand the set of primary information received by the device, and investigate the nature of seawater anomalies with their quantitative and qualitative assessment.

The advantage of the proposed modified device with two acoustic measuring channels is that, on the basis of existing SVP profilers, without any special design changes, it is possible to obtain a meter with new functionality.

As a further development of the modified device, it is proposed to expand the operating frequency range to 10 MHz and supplement it with one more measuring channel for monitoring ultrasound scattering, which will make it possible to investigate the attenuation components and separate the absorption separately.

**REFERENCES**

1. *Zhou, Qifa, Sienting Lau, Dawei Wu, and K. Kirk Shung.* "Piezoelectric films for high frequency ultrasonic transducers in biomedical applications." Progress in materials science 56, no. 2 (2011): 139–174.

2. *Fisher F.H., Simmons V.P.* Sound absorption in sea water // J. Acoust. Soc. Am. **62**, 558–564 (1977).

3. *Mellen R.H., Simmons V.P., Browning D.G.* Sound absorption in sea water: A third chemical relaxation // J. Acoust. Soc. Am. **65**, 923–925 (1979).

4. *Fisher F.H., Simmons V.P.* Sound absorption by a third chemical reaction // J. Acoust. Soc. Am. **65**, 1327–1329 (1979).

5. *Kibblewhite A.C., Hampton L.D.* A review of deep ocean sound attenuation data at very low frequencies // J. Acoust. Soc. Am. **67**, 147–157 (1980).

6. *Pinkerton M.M.* A pulse method for the measurement of ultrasonic Absorption in liquids: results for water // Nature. **160**, 128 (1947).

7. *Leonard R.W.* The attenuation of ultrasonic waves in water // J. Acoust. Soc. Am. **20**, 224 (1948).

8. *Liebermann L.* The origin of sound absorption in water and in sea water // J. Acoust. Soc. Am. **20**, 868 (1948).

9. *Francois R.E., Garrison G.R.* Sound absorption based on ocean measurements. Part I: Pure water and magnesium sulfate contributions // J. Acoust. Soc. Am. **72**, 896–907 (1982).

10. *Francois R.E., Garrison G.R.* Sound absorption based on ocean measurements. Part II: Boric acid contribution and equation for total absorption // J. Acoust. Soc. Am. **72,** 1879–1890 (1982).

11. *Browning D.G., Mellen R.H.* Attenuation of low-frequency sound in the sea: Recent results, in *Progress in Underwater Acoustics*, edited by H.M. Merklinger (Plenum, New York, 1987). P. 403–410.

12. *Skretting A., Leroy C.C.* Sound Attenuation between 200 Hz and 10 kHz // J. Acoust. Soc. Am. **49**, 276–282 (1970)

13. *Mellen R.H., Browning D.G.* Variability of Low-Frequency Sound Absorption in the Ocean: pH Dependence // J. Acoust. Soc. Am. **61**, 704–706 (1977).

14. *Garrison G.R., Early E.W., Wen T.* Additional Sound Absorption Measurements in near-Freezing Sea Water // J. Acoust. Soc. Am. **59**, 1278–1283 (1976).

15. *Grekov A.N., Grekov N.A., Sychov E.N.* The use of sound velocity profilers to determine the density of water. The Proceedings of the First International Conference on Ocean Thermohydomechanics ICOT-2017. 28–30 November 2017 in Moscow in the Shirshov Institute of Oceanology, Russian Academy of Sciences. 2017. P. 46–49.

16. *Grekov A.N., Grekov N.A., Shishkin Y.E.* Investigation of a sound speed profilograph characteristics and correction of measurement results // Monitoring systems of environment. 2017. Vol. 10 (30). P. 24–30. https://doi.org/10.33075/2220-5861-2017-4-24-30

17. *IOC,* SCOR and I A PSO, 2010: The international thermodynamic equation of seawater – 2010: Calculation and use of thermodynamic properties. Intergovernmental Oceanographic

Commission, Manuals and Guides No. 56, UNESCO (English), 196 p., available at: http://www.TEOS-10.org

18. *Del Grosso V.A., Mader C.W.* Speed of sound in pure water // J. Acoust. Soc. Amer. 1972. Vol. 52. N 5. P. 1442–1446.

19. Belogol'Skii V.A. et al. Pressure dependence of the sound velocity in distilled water // Measurement Techniques. 1999. Vol. 42 (4). P. 406-413.

20. Ultrasonic-Flow-Converter Data Sheet TDC-GP22 March 13th 2014 Document-No: DB_GP22_en V0.9 Universal 2-Channel Time-to-Digital Converters Dedicated to Ultrasonic Heat & Water Meters (Available at https://ams.com/documents/20143/36005/TDC-GP22_DS000323_1-00.pdf/aa6c41ca-1312-60ec-f6a8-82c90da3f856).

21. *Jobst S., Rudolf B*. A Comparison of Correlation and Zero-Crossing Based Techniques in Ultrasonic Measurements // In Proceedings of "Applied Research Conference 2014", Ingolstadt, 2014. P. 362–267.

22. Ultrasonic-Flow-Converter Data Sheet TDC-GP30 June 27th, 2019 Document-No: DB_GP30Y_Vol1_en V0.3 System-Integrated Solution for Ultrasonic Flow Meters Volume 1: General Data and Frontend Description (Available at https://ams.com/documents/20143/36005/TDC-GP30_DS000391_3-00.pdf/f96f8c8b-87e5-ac8d-a26c-65756fd240fa)

23. *Grekov A.N., Grekov N.A., Sychov E.N.* New Equations for Sea Water Density Calculation Based on Measurements of the Sound Speed // Mekhatronika, Avtomatizatsiya, Upravlenie. 2019. 20 (3). P. 143–151. https://doi.org/10.17587/mau.20.143-151.